# Investigation of Microstructure and Corrosion Resistance of Ti-Al-V Titanium Alloys Obtained by Spark Plasma Sintering

Aleksey Nokhrin, Pavel Andreev, Maksim Boldin, Vladimir Chuvil'deev, Mikhail Chegurov, Ksenia Smetanina, Mikhail Gryaznov, Sergey Shotin, Artem Nazarov, Gleb Shcherbak, Artem Murashov and Galina Nagicheva

*Lobachevsky State University of Nizhniy Novgorod*

Correspondence: andreev@phys.unn.ru

**Abstract:** The research results of the microstructure and corrosion resistance of Ti and Ti-Al-V Russian industrial titanium alloys obtained by spark plasma sintering (SPS) are described. Investigations of the microstructure, phase composition, hardness, tensile strength, electrochemical corrosion resistance and hot salt corrosion of Ti-Al-V titanium alloy specimens were carried out. It was shown that the alloy specimens have a uniform highly dense microstructure and high hardness values. The studied alloys also have high resistance to electrochemical corrosion during tests in acidic aqueous solution causing the intergranular corrosion as well as high resistance to the hot salt corrosion. The assumption that the high hardness of the alloys as well as the differences in the corrosion resistance of the central and lateral parts of the specimens are due to the diffusion of carbon from the graphite mold into the specimen surface was suggested.

## 1. Introduction

At present, titanium (Ti) alloys are one of the principal materials in nuclear and marine engineering [1–4]. α- and near-α Ti alloys (Russian industrial names PT-3V, PT-7M, etc.) are used most often in nuclear power engineering [3,4], new two-phase α + β alloys are increasingly used in marine engineering in recent years [1,2]. The extreme operational conditions (high temperatures, loads, impact of corrosion-aggressive ambient, radiation environment, etc.) to which they are exposed impose high requirements of the mechanical properties and corrosion resistance of the Ti alloys. An increase in the strength and corrosion resistance of the Ti alloys will allow to produce lighter and more reliable products for marine mechanical engineering, aircraft building, etc.

Hot salt corrosion (HSC) is one of the most dangerous kinds of destructive corrosion for Ti products utilized at elevated temperatures and in corrosion-aggressive media [5–7]. HSC develops intensively beneath the deposits of crystal halogenide salts (chlorides, bromides, iodides) in the presence of water (bound in the salt crystals, occluded in the salt crystals, or present in ambient air) on the Ti alloy surfaces. The character of HSC (general corrosion, pitting corrosion, or intergranular corrosion) depends on the operational conditions (atmospheric humidity, oxygen pressure, temperature, thickness and composition of the salt deposits) and on the structure and phase composition of the Ti alloys [5]. Crystalline chlorides cause the most intensive HSC of alkaline and alkaline earth metals. The case of the intergranular HSC, which may lead to formation of cracks and rapid destruction of the Ti alloys in the external stress conditions, is the most dangerous [5,8,9].

Various methods of deformation treatment (e.g., various severe plastic deformation (SPD) methods) are often applied to improve strength and corrosion resistance of the Ti alloys [10–12]. In recent years, modification of the chemical composition of the Ti alloys (e.g., by doping with Pt-group metals) are also applied to improve strength and corrosion resistance of the ones [13,14]. In spite of some success in this area, it is worth mentioning that the possibilities of the expensive doping of the Ti alloys are often limited by economic feasibility. Besides, the doping of Ti with the Pt-group metals (Ru, Pt, etc.) doesn't lead to essential improvement of strength often [13,14]. The prospects of application of SPD methods are often limited by possibilities of application of ultrafine-grained (UFG) materials for long-term operation at elevated temperatures. In particular, it was shown that a long-term HSC test (250 °C, 500 h) of the UFG near-α Ti alloy specimens led to the development of recrystallization and softening of the alloys [15]. Note also that the change in chemical and phase composition of the grain boundaries in the UFG near-α Ti alloys is possible during recrystallization. It may lead to a change in the HSC character in the Ti alloys-from the general and/or pitting corrosion at short test duration to the most dangerous intergranular corrosion at long testing times [15]. Therefore, the capabilities of modern powder metallurgy methods, which may allow the formation of the high-strength thermally stable state of the Ti alloys, are interesting.

Spark plasma sintering (SPS) is one of the most promising methods of controlling the structure and phase composition of metallic materials [16,17]. In this method, the sintered material is heated up rapidly (up to 2500 °C/min) by passing millisecond pulses of high-power electric current through the mold with the powder with simultaneous application of pressure. Materials obtained by SPS feature high density (close to the theoretical one) and mechanical properties [18]. As one can see from Table 1, which summarizes the work results [19–38], the Ti alloys obtained by SPS have high hardness and strength as well as good plasticity (for certain sintering conditions). In some cases, the Ti alloys obtained by SPS have high corrosion resistance [33–35]. At present, SPS has found its application as a method of high-speed solid-phase diffusion welding or bonding of Ti alloys [39,40] as well as an efficient method of fabrication of Ti-based dispersion-hardened composites (Ti-$TiB_2$, Ti-$Si_3N_4$, Ti-$ZrO_2$, etc.) with increased strength [22,26,35,41,42]. At the same time, it should be noted that corrosion resistance of the Ti alloys obtained by SPS remains almost unexplored to date. A special interest attracts studying the resistance of the Ti alloys to hot salt corrosion, which is one of the most dangerous kinds of damaging processes (see above).

The purpose of the work is to study the microstructure, mechanical properties and corrosion resistance of the Ti alloys obtained by SPS method.

**Table 1.** Microstructure parameters and mechanical properties (hardness, tension testing results) of some Ti alloys obtained by SPS.

| Alloy | Initial Powders | SPS Mode | | | | Material Characteristics | | | | | | Ref. |
|---|---|---|---|---|---|---|---|---|---|---|---|---|
| | | $V_h$ (°C/min) | T (°C) | P (MPa) | t (min) | XRD Analysis | ρ (%) | d (μm) | $Hv$ | $σ_b$ (MPa) | δ (%) | |
| Ti | $R_0$< 75 μm, purity >98% | 50 | 1200 1350 | 50 | 5 | α + β | 99.8 ~98 | - | ~391 ~360 | ~500 ~390 | ~4.2 ~2.5 | [19] |
| Ti | Grade 1 Grade 3 | - | 900 | 60 | 5 | - - | ~100 ~100 | ~15 ~12 | ~145 ~240 | ~450 ~700 | 38–39 ~18 | [20] |
| Ti | Grade 2 + CM [2] (8 h) | 50 50 100 50 | 900 850 850 850 | 80 80 80 40 | 5 3 3 3 | - | 99.16 99.2 99.2 98.48 | - - - - | ~244 ~247 ~247 ~240 | 776 838 823 758 | 29.3 26.4 19 18.4 | [21] |
| Ti | - | - | 1050 | 50 | 5 | α-Ti | 97.9 | 40–50 | 291 | - | - | [22] |
| **Table 1.** Cont. | | | | | | | | | | | | |
| Ti | Grade 4 (20–60 μm)+ CM [2] (8 h) | 100 | 800 | 80 | 3 | - | ~99 | ~4.2 | ~260 | - | - | [23] |
| Ti | 99.8% (HEBM [2]) | 100 | 1200 | 50 | 10 | - | ~97 | | 341 | - | - | [24] |
| Ti-6Al-4V | HEBM [2] 0 h 10 h 30 h | 100 | 850 | 50 | 4 | - α + β | ~100 87.4 99.3 - | ~15 ~0.87 ~0.58 | ~370 - - - | - ~1200 1663 1735 | - ~40 20 10 | [25] |
| Ti-6Al-4V | $R_0$ ~ 13.4 μm | 293 | 900 | 70 | 10 | - | ~97.5 | 10–20 | ~4 | ~1050 | ~10 | [26] |
| Ti-6Al-4V | purity 99.9% | 200 300 400 | 1000 1000 1000 | 5 25 50 5 25 50 5 25 50 | 5 5 5 | α + β | 94.58 98.87 98.65 93.45 100.0 97.74 96.16 99.55 99.32 | - 120.44 110.19 115.49 143.57 86.94 - 118.23 84.31 | 293 346 348 325 352 373 318 345 352 | 1169 [1] 1281 [1] 1310 [1] 1287 [1] 1406 [1] 1394 [1] 1310 [1] 1414 [1] 1394 [1] | - | [27,28] |
| Ti-6Al-4V | Pure Ti-6-4 HEBM [2] +1% MWCNT [2] HEBM [2] +2% MWCNT [2] HEBM [2] +3% MWCNT [2] | 100 | 1000 | 50 | 5 | α + β α + β + CNT + TiC | ~99.6 ~99 ~98.2 ~98 | - - - - | ~360 ~380 ~395 ~400 | - - - - | - - - - | [29] |
| Ti-6Al-4V | $R_0$ ~ 25 μm, purity >99.9% | 100 | 1000 | 50 | 10 | α + β | 97–99 | - | 389 | - | - | [30] |
| Ti-6Al-4V | $R_0$ ~ 65 μm | 100 | 1250 | 50 | 15 | α + β | 99.8 | - | 450 | 1057 | 15 | [31] |
| Ti-6Al-4V | $R_0$ ~ 25 μm, purity >99.9% | 100 | 1000 | 50 | 10 | α + β | 99.54 | - | 389 | - | - | [32] |
| Ti-6Al-4V | $R_0$ ~ 45–90 μm | 100 | 1000 | 50 | 6 | α + β | 98.6 | - | 325.46 | - | - | [33,34] |
| Ti-6Al-4V | $R_0$ ~ 60 μm, purity >99% | 100 | 1000 | 30 | 10 | α + β | 98.22 | - | 3.25 GPa | 1060 | - | [35] |
| Ti-6Al-4V | - | - | 650 | 555 | - | α-Ti | 98.9 | 11.7 | 3.7 GPa | - | - | [36] |
| Ti-6Al-4V | $R_0$ ~ 75–150 μm | 100 | 1050 | 30 | 5 | α + β | - | - | - | 1240 | 19.5 | [37] |
| Ti-6Al-4V | $R_0$ ~ 53–106 μm | 100 | 900 | 50 | 5 | α + β | - | - | - | 844 | 12 | [38] |

[1] - Compression test; [2] - Powder technology: CM - Cryomilling, HEBM - High energy ball milling; MWCNT - multi-walled carbon nanotubes

## 2. Materials and Methods

The objects of this study were a Ti α-alloy Russian industrial name VT1-0, near-α Ti alloy PT-3V (Ti-5wt.%Al-2wt.%V, β-phase particles' content no more than 5%) and the two-phase α + β alloy VT-6 (composition: Ti-6wt.%Al-4wt.%V). The exact chemical compositions of the alloys are presented in Table 2. The chemical composition of the alloys corresponds to Russian Standard GOST 19807–91. The alloys were sintered from powders manufactured by plasma atomization in an inert gas ambienceby Normin Co. (Borovichi, Russia). According to the vendor's certificate, the powders belonged to PNT-8 class (the particles have a spherical shape according to Russian National Standard GOST 25849–83 with sizes ranging from 10 to 45 μm). The bulk density of the powders was 2.7 g/cm$^3$.

**Table 2.** Chemical composition of the initial powders (vendor's certificate data).

| Alloys | Elements, wt.% | | | | | | | | | |
|---|---|---|---|---|---|---|---|---|---|---|
| | Ti | Al | V | Zr | Si | Fe | O | H | N | C |
| Ti (VT1-0) | base | - | - | - | 0.01 | 0.14 | 0.13 | 0.006 | 0.038 | 0.021 |
| Ti5Al2V (PT-3V) | base | 4.4 | 1.95 | 0.15 | 0.02 | 0.06 | 0.13 | 0.0049 | 0.039 | 0.015 |
| Ti6Al4V (VT-6) | base | 6.24 | 4.05 | - | - | 0.12 | 0.12 | 0.006 | 0.020 | 0.016 |

SPS of Ti specimens of 20 mm in diameter was performed on a Dr. Sinter model SPS-625 instrument (SPS Syntex Inc., Tokyo, Japan). The heating rate was 50 °C/min. The Ti powders were placed into a graphite mold, the interior of which was protected by two layers of graphite paper (graflex). The BN protective coating was not applied. Sintering was carried out in a vacuum at constant uniaxial pressure of 70 MPa. The sintering temperature (Ts) was 850 ± 10 °C that corresponds to the two-phase (α + β)-region [2,43]. The temperature was measured by an IR-AHS2 optical pyrometer (Chino, Torrance, CA, USA) focused on the surface of the graphite mold. The shrinkage of the powders was monitored by a built-in Dr. Sinter model SPS-625 dilatometer. After sintering, the specimens' surfaces were subjected to mechanical grinding and polishing up to the roughness of 3–5 μm in order to remove residual graphite contaminations.

The density of the specimens was measured by hydrostatic weighting (Archimedes method) using a CPA 225D balance(Sartorius, Göttingen, Germany). When calculating the relative density ($\rho$) the theoretical density was taken to be $\rho_{th}$ = 4.5 g/cm$^3$.

The microstructural investigations were carried out using an IM DRM metallographic optical microscope (Leica Microsystems, Wetzlar, Germany) and a JSM-6490 scanning electron microscope (SEM, JEOL, Tokyo, Japan) with an INCA 350 energy dispersion (EDS, Oxford Instruments, Abingdon, UK) analyzer. The alloy specimens were subjected to mechanical polishing and electrolytic etching in 75% $H_2SO_4$ + 15% $HNO_3$ + 10% HF solution at room temperature to reveal the microstructure. The X-ray diffraction (XRD) phase analysis was carried out using an XRD-7000 X-ray diffractometer (Shimadzu, Kyoto, Japan, CuKα-filter, $\lambda_{Cu}$=1.54078Å, scanning range 2θ = 30–80°, scanning step 0.02°, exposure 2 s, "wide slit" mode. The technique of investigating the Ti alloys was described in [44]. The crystal structure data for the phase composition analysis were taken from Powder Diffraction File-2 (PDF-2) of the International Centre for Diffraction Data (ICDD) and the Inorganic Crystal Structure Database (ICSD). The evaluation of the phase composition of the Ti alloys was performed by the reference intensity ratio (RIR) method. Calculations of internal strain in the sintered specimens were performed using the Williamson-Hall model [45]. The unit cell parameters of Ti were refined by the Rietveld method.

The Vickers hardness ($H_v$) measurements were performed using a Qness A60+ hardness tester (ATM Qness, Golling an der Salzach, Austria) with a 2 kg load. The tension tests were carried out at room temperature using a Tinius Olsen H25K-S tension machine (Tinius Olsen, Horsham, PA, USA) at the deformation rate of 0.01 mm/s. Flat 25 mm long and 2 thick "double-blade" shaped specimens were cut out for the testing from the bulks of 30 mm in diameter by electric spark cutting. The correspondence between the structure and properties of the specimens of 20 mm and 30 mm in diameters were confirmed by the results of density and microhardness measurements as well as by the metallographic and SEM investigations. The fractographic analysis of the fractures was carried out using the Jeol JSM-6490 SEM.

The electrochemical investigations of corrosion resistance were carried out using anR8 potentiostat-galvanostat (Electrochemical Instruments JSC, Chernogolovka, Russia) at room temperature in aqueous 0.2%HF + 10%$HNO_3$ solution (pH = 1.1) causing the intergranular corrosion in Ti alloys (see [15]). Testing was carried out in a glass electrochemical cell of 60 ml volume, in a still solution. The temperature of testing was 30°C. A chlorine-silver reference electrode (EVL-1M4) and a Pt wire as an auxiliary electrode were used in testing. The specimens' surfaces were subjected to mechanical grinding and polishing (with diamond paste 3/5) up to the roughness of 3–5 μm. The final polishing stage was done manually to minimize surface hardening. Prior to conducting the electrochemical testing, the specimens' surfaces were covered by a corrosion-proof compound except open (corrosion-active) areas of 0.3 $cm^2$ in the central parts of the specimens (Zone 1) and edges of the specimens (Zone 2).

During the first stages of the electrochemical studies, the potentiometric curves "potential $E$-current $i$" were measured in the range of potentials from –800 mV to +500 mV with the potential sweep rate of 0.5 mV/s. Then, prior to recording Tafel curves the specimens were exposed to the electrolyte for 1 h with simultaneous measurement of the $E(t)$ dependence. At the third stage of investigations, the corrosion current density $i_{cor}$ and the corrosion potential $E_{cor}$ were determined from analysis of the Tafel parts of the "potential-current density" curves in semi-logarithmic axes lg$i$–E. Besides, the slopes of the anodic parts of the Tafel regions in the dependencies lg$i$-E (*A*) and the cathodic ones (*C*) were determined. The sloes *A* and *C* were determined using the Electro Chemical Instruments ES8 software unit included with the R8 potentiostat-galvanostat control software. The lg$i$-E curves were recorded with the potential sweep rate 0.5 mV/s.

The autoclave testing for hot salt corrosion was performed in a mixture of NaCl and KBr (300:1) at 250 °C according to the technique described in [46] in the conditions of oxygen access. The testing time was 500 h. The degree of the corrosion damage was evaluated according to Russian National Standard GOST 9.908-85. The analysis of depth and character of the corrosion defects was performed using the Leica DM IRM metallographic optical microscope.

## 3. Results

### 3.1. Microstructure and Mechanical Properties Investigation

The particles of all initial powders had a spherical shape (Figures 1a,b) and close granulometric composition. The particle sizes ranged from 10 to 40 μm that corresponds to the vendor's certificate data.

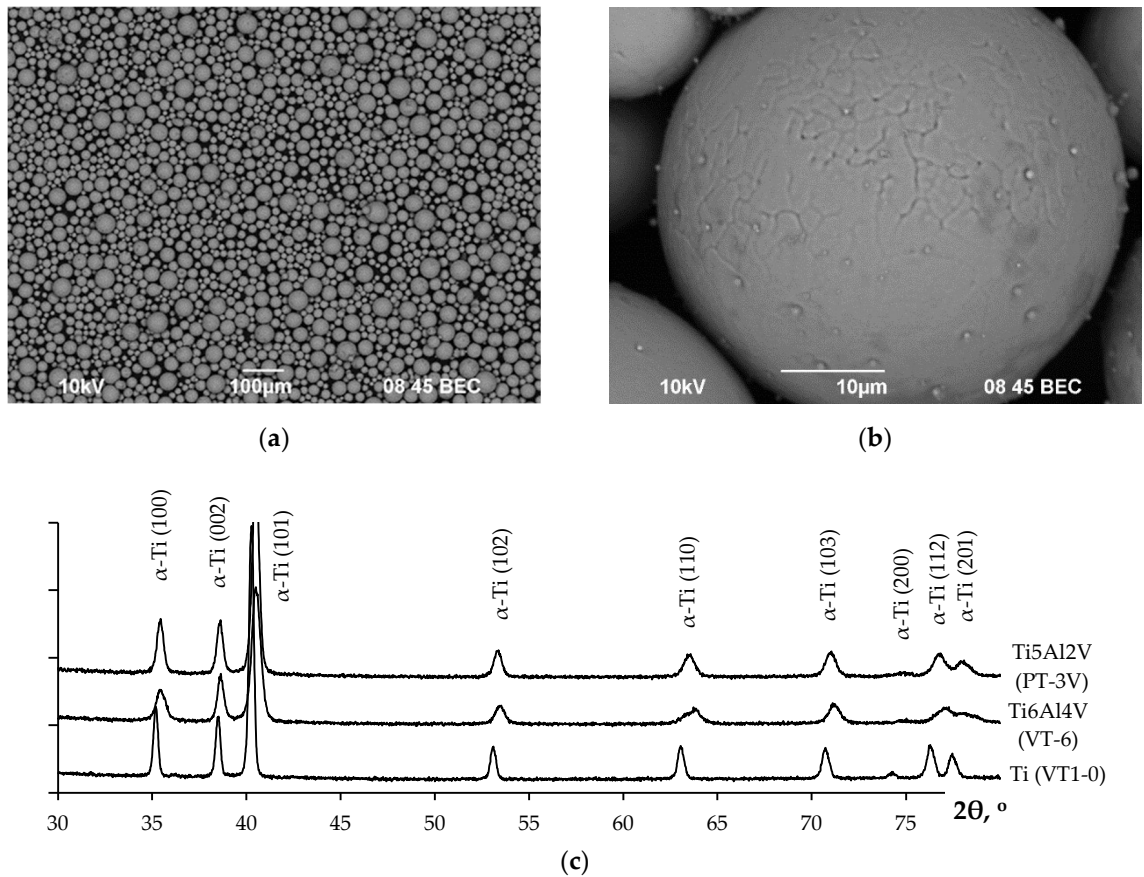

**Figure 1.** SEM images (BEC mode) of Ti5Al2V powder (**a**,**b**) and XRD spectra (**c**) of initial powders.

The summary of the XRD phase analysis in Figure 1c shows the peaks corresponding to the α-Ti phase only to be present in the XRD patterns of the powders. The diffraction peak positions on the XRD patterns are close to the theoretical values for α-Ti. It allows concluding essential crystal lattice distortions to be absent in the initial powders.

The SEM images of the microstructure of the sintered specimens are presented in Figure 2. It can be seen that the alloys had a dense microstructure with well-defined plate-shaped grains of the α-phase. The mean size (length) of the α-phase plates ($d_\alpha$) in the Ti (VT1-0) material was 50–70 μm, the ones in the Ti5Al2V and Ti6Al4V alloys were 10–15 μm. So far, the average grain size in the sintered specimens were close to the ones of the initial particles. Note that more uniform equiaxial microstructure was formed in the Ti5Al2V alloy than in the Ti6Al4V. The boundaries of the α-phase grains in the Ti6Al4V alloy had a lighter contrast in Z-contrast mode that evidences the nucleation of β-phase particles at the α-phase grain boundaries (Figures 2d and 3, see [39,46,47]). Similar "light" grain boundaries were found in the Ti5Al2V alloy as well although the volume fraction of these ones was lower than in the Ti6Al4V alloy considerably (Figure 3).

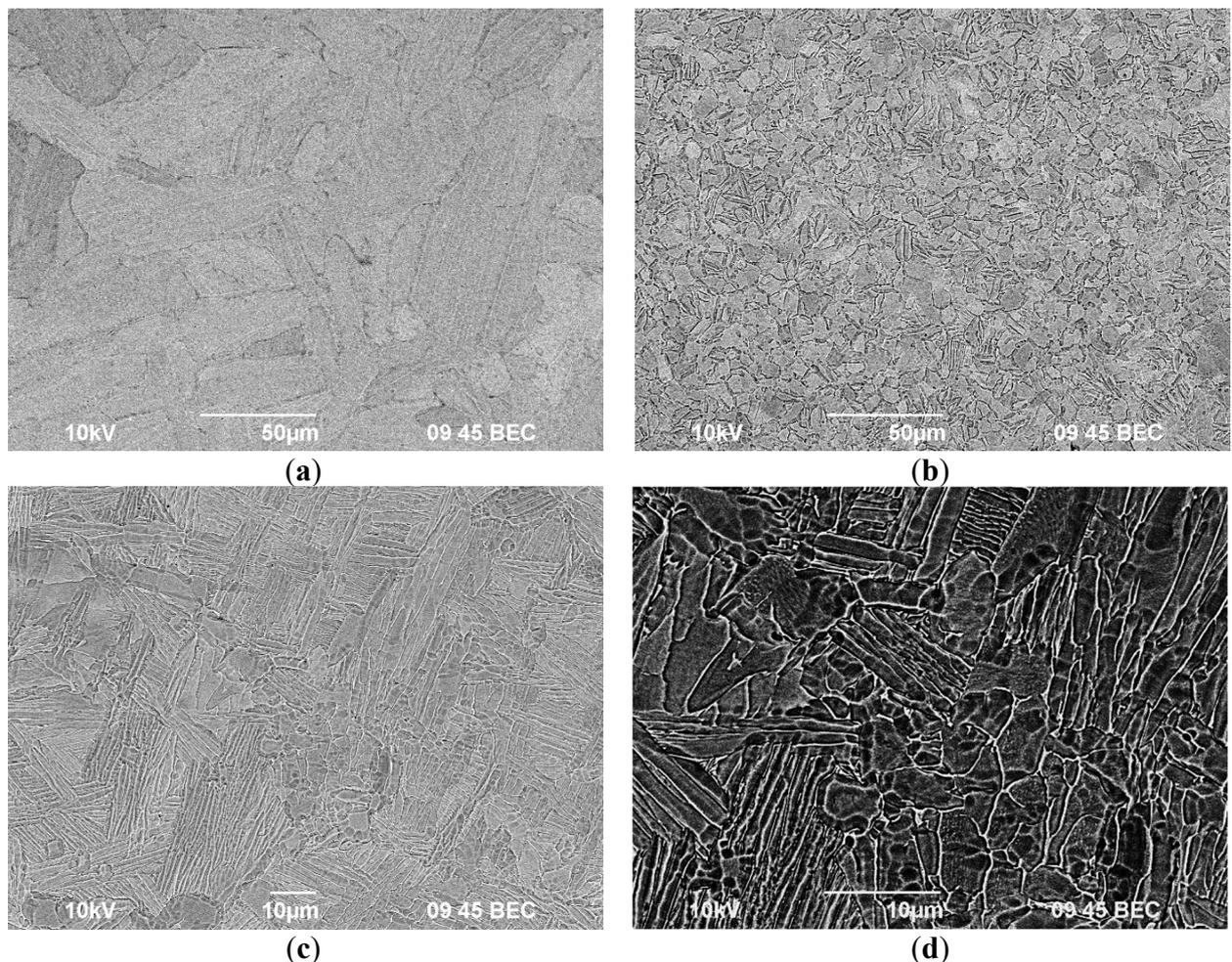

Figure 2. Microstructure of the Ti (a), Ti5Al2V (b), and Ti6Al4V (c,d) alloys. SEM (BEC mode)

The densities of the Ti alloys were high enough. For the Ti, Ti5Al2V, and Ti6Al4V alloys, these ones were 99.77% (4.49 g/cm$^3$), 97.63% (4.393 g/cm$^3$), and 98.42% (4.429 g/cm$^3$), respectively. Note that no large pores were found in the microstructure of the alloys.

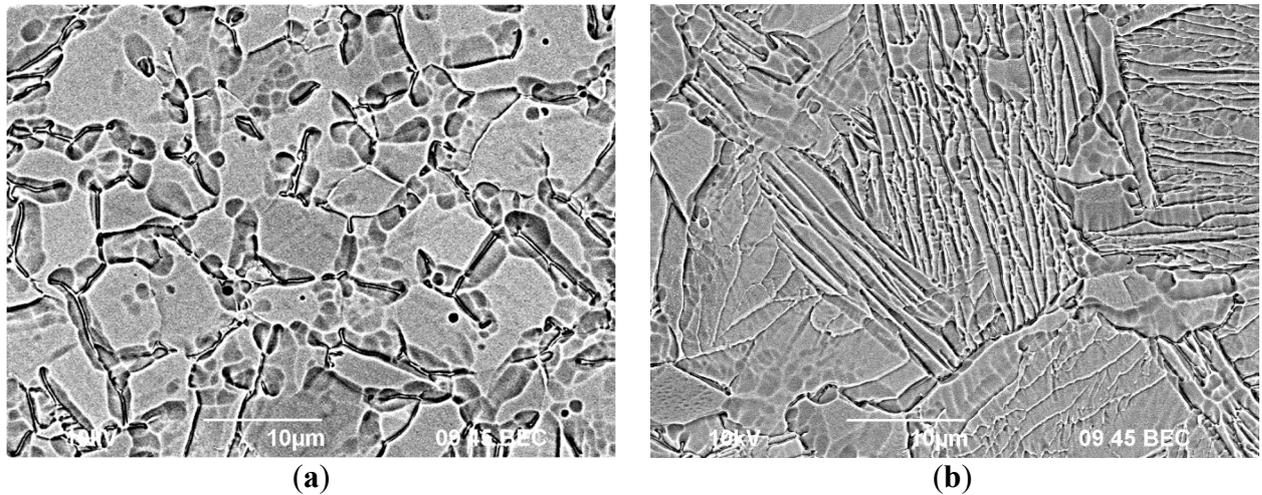

(a)      (b)

**Figure 3.** Microstructure of the grain boundaries with β-phase particles in the Ti5Al2V (**a**) and Ti6Al4V (**b**) titanium alloys. SEM (BEC mode)

Analysis of the XRD results presented in Figure 4 has shown that only peaks corresponding to the α-phase (PDF #00-044-1294, ICSD #4390) are observed in the Ti (VT1-0 alloy) specimen. Two phases were found in the Ti6Al4V and Ti5Al2V specimens. The structures of these two matched to the α-Ti and β-Ti (PDF #01-074-7075, ICSD #44391) phases. No other crystalline phases were found. Quantitative phase analysis carried out by the RIR method has shown that the mass fraction of the β-phase particles in the Ti5Al2V and Ti6Al4V alloys to be ~1.5% and about 5%, respectively. No XRD peaks corresponding to nucleation of other phases (e.g., α' or α"), which may nucleate in these alloys at high temperature plastic deformation [48,49], were observed. The unit cell parameters of α-Ti in the sintered specimens were close to the ones in the initial powders.

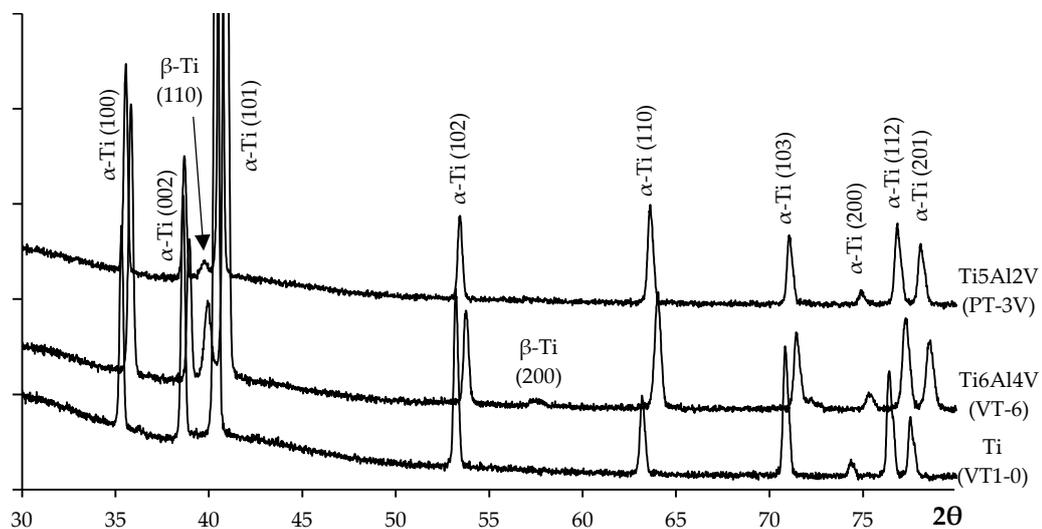

**Figure 4.** XRD spectra of the sintered Ti alloys.

The average values of hardness of the Ti, Ti5Al2V and Ti6Al4V alloys were 1.8–1.9 GPa, 3.3–3.4 GPa and about 3.2 GPa, respectively. The values of hardness of the central parts of the specimens (Zone 1) were lower than the ones on the edges of the specimens adjacent to the graphite mold (Zone 2) insufficiently (by ~0.1 GPa). All the imprints of the diamond pyramidal indenter had a regular shape with absence of cracks.

It should be noted that the Ti alloys obtained by SPS have rather high hardness (1.3–1.5 times higher than the hardness of the coarse-grained alloys with similar composition obtained by traditional technology of hot deformation in the β-region). In particular, hardness of the Ti5Al2V alloy obtained by SPS (3.3–3.4 GPa) was considerably higher than the one of the coarse-grained Ti5Al2V alloy with the grain sizes 10–50 μm (~2.0–2.1 GPa [39,46]). It is a quite unexpected result since the microstructure parameters of the Ti alloys obtained by SPS and by traditional hot deformation method are close enough. As one can see from Table 1, the high values of hardness are typical for the Ti alloys obtained by SPS.

High hardness of the Ti alloys cannot be related to the internal stresses, which may appear in the specimens as a result of the plastic deformation in the sintering process. The analysis of the XRD results using Williamson-Hall model has shown the magnitude of the internal microstress to be negligible.

High mechanical properties of the Ti alloy specimens were confirmed by the results of tension mechanical tests. The tension curves stress ($\sigma$)-strain ($\varepsilon$) are presented in Figure 5. One can see the tensile strength ($\sigma_b$) of the specimens is high. In particular, for the Ti5Al2V alloy it was ~1160–1170 MPa. The value of $\sigma_b$ is close to the tensile strength of the Ti5Al2V alloy after equal channel angular pressing (~1050 MPa) [46]. The magnitude of tensile strength of the Ti and Ti6Al4V alloy obtained by SPS were also very high and reached 570–580 MPa and 1030–1040 MPa, respectively. It is important to mention that the Ti alloys obtained by SPS have very good plasticity. As one can see in Figure 5, the stages of stable (uniform) plastic flow were observed in the tension curves. The magnitude of the ultimate elongation to failure in the Ti, Ti5Al2V and Ti6Al4V alloys were 34–35%, 14–15%, and 25–26%, respectively. These are high enough characteristics pointing to the prospects of forming a state with simultaneously increased strength and plasticity in these alloys by SPS. The comparison of the results of mechanical tests with the data presented in Table 1 shows that the values of tensile strength obtained for the Ti5Al2V alloy to be the record (Ti5Al2V alloy is a Russian analog of Grade 9 but differs in an increased Al content).

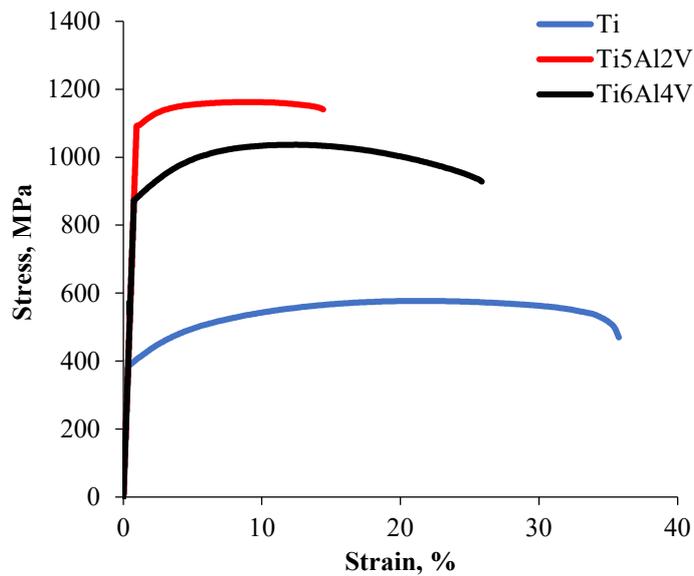

**Figure 5.** Stress-strain curves for the Ti alloy specimens obtained by SPS.

The results of the fractographic analysis of the Ti alloy specimen fractures after the tension tests at room temperature are presented in Figure 6. The specimen fractures were uniform and no fracture defects were observed. The fracture microstructure was classified according to [50]. One can distinguish the zones of arising and propagation of the cracks and the failure zones on the fracture surfaces of the Ti specimens. Respective zones are marked as "1" and "2" in Figures 6b,e. The destruction areas along the slip planes occupy almost the whole upper part of the fracture of the VT1-0 specimen (Figure 6c). The failure zone is a set of elongated pits of various sizes (Figure 6d). The structure of the fractures of the Ti6Al4V specimens had the same characteristics areas as the fractures of the Ti specimens VT1-0 (Figures 6e–g).

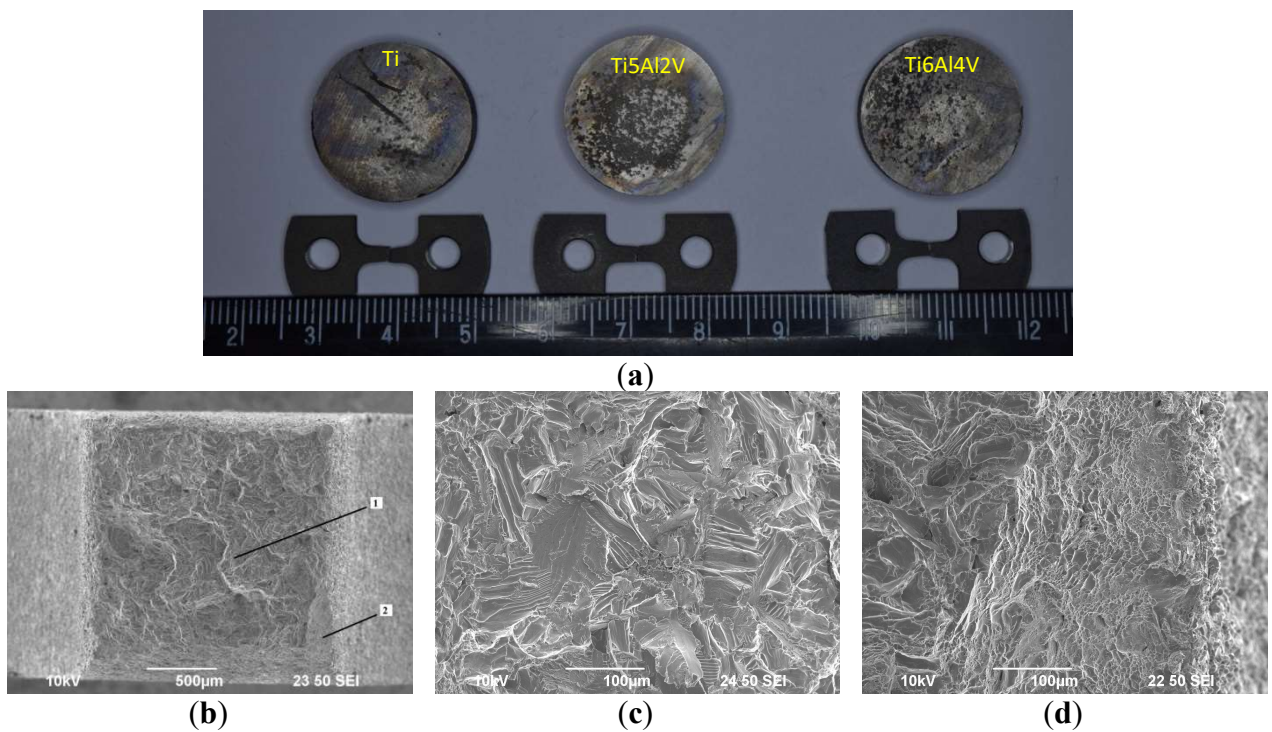

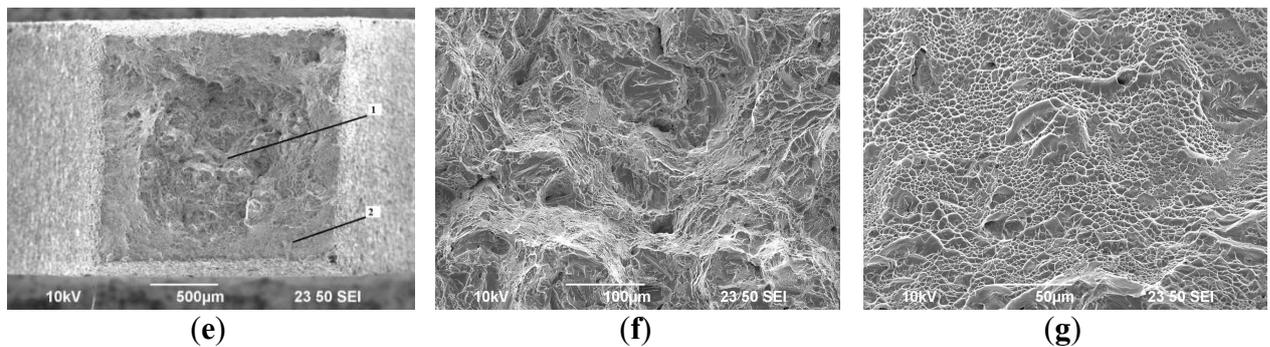

| (e) | (f) | (g) |

**Figure 6.** General view of the specimens after tension tests (**a**) and the fractographic analysis of the fractures of the Ti alloy (**a**), (**b,e**) - general view of fractures, zones of arising and propagation of the cracks and the failure zones are marked as "1" and "2", respectively, (**c,f**) - regions of arising and propagation of the cracks, (**d,g**) - failure zones (SEM).

The fractures of the Ti5Al2V alloy specimens had more complex character (Figure 7). The fracture surfaces were heterogeneous in macro geometry; one can distinguish the destruction zones. No destruction foci were found. The zone of arising and slow propagation of the crack (marked by "1" in Figure 7a, an enlarged image is shown in Figure 7b), the zone of rapid propagation of the crack (marked by "2" in Figure 7a, an enlarged image is shown in Figure 7c), and the failure zone (marked by "3" in Figure 7a, an enlarged image is shown Figure 7d) as well as defects of round or ellipsoidal shapes and about 100–150 μm in sizes (marked by "4" in Figure 7a, the enlarged images are shown in Figures 7e,f) were observed on the fracture. The zone of arising and propagation of the crack is a set of shallow pits that evidences a small degree of tensile deformation indirectly. A brittle destruction was observed in the defect area consisting of a chip and chip facets (Figures 7e,f). In our opinion, the presence of single large oxide particles in the initial powder may be one of the formation origins of the defects with characteristic round shapes.

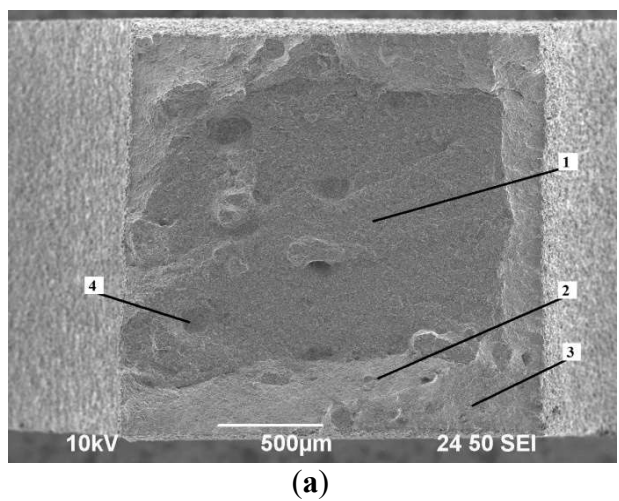 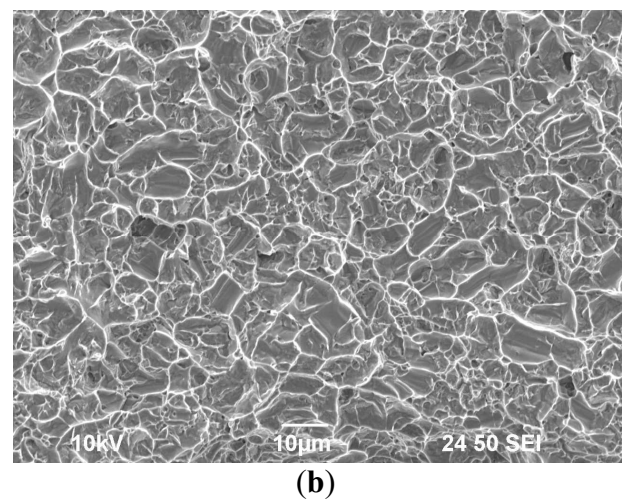

| (a) | (b) |

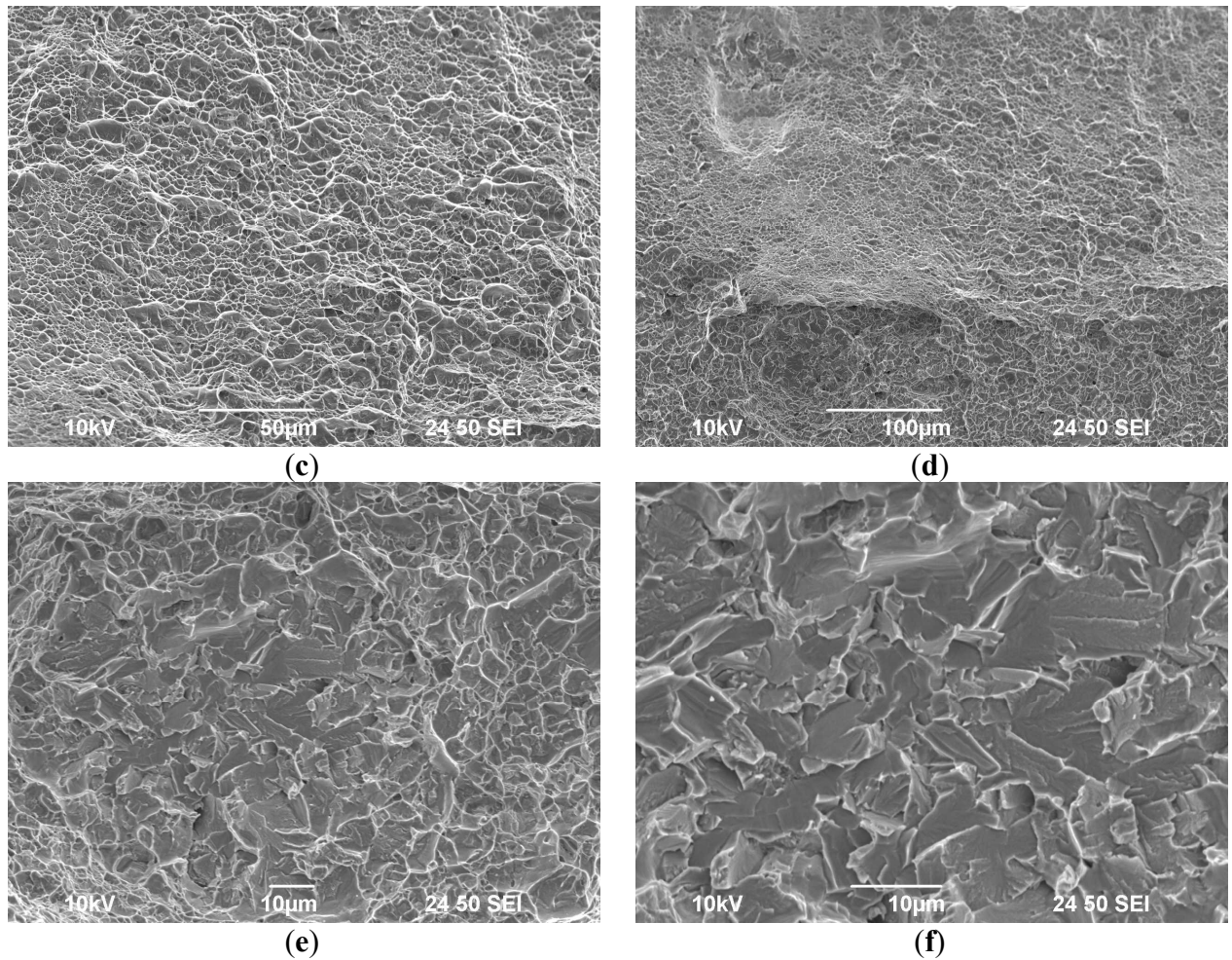

(c)  (d)

(e)  (f)

**Figure 7.** SEM images of the fracture zones analysis of the fractures of the Ti5Al2V alloy specimens after the tension tests: (**a**) general view of a fracture; (**b**) zone of arising and slow propagation of a crack marked by "1" in (a); (**c**) zone of rapid propagation of the cracks marked by "2" in (a); (**d**) failure zone marked by "3" in (a); (**e**,**f**) fractographic of defect fracture marked by "4" in (a).

*3.2. Corrosion Resistance Investigations*

After the HSC testing, dense dark films with some light inclusions of NaCl were observed on the surfaces of the Ti specimens (Figure 8a). After rinsing in hot water, leading to the dissolving of the salt, thick dark films with some pores distributed uniformly were observed on the specimens' surfaces (Figure 8b). The presence of pores in the films evidences the HSC process proceeds with a release of the gas-phase $TiCl_4$ (see [5]).

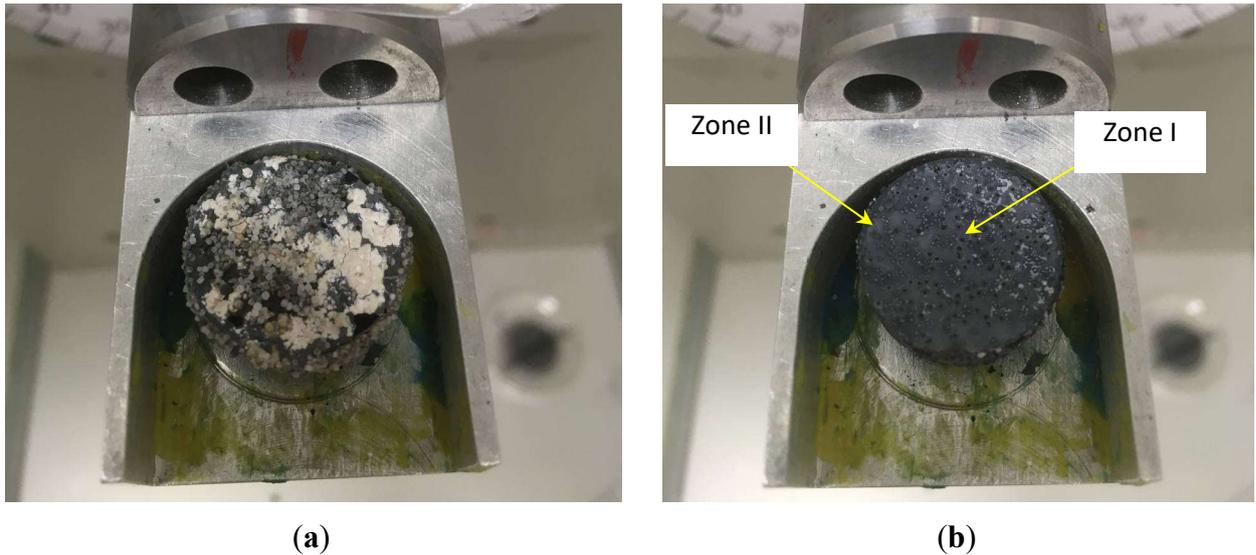

**Figure 8.** General view of the Ti (**a**) and Ti5Al2V (**b**) alloys' specimens after HSC testing (**a**) and after rinsing in hot water (**b**)

Summarizing the XRD phase analysis results presented in Figure 9, the oxides $Na_4Ti_5O_{12}$ (PDF #00-052-1814, ICSD #170677), $TiO_2$ (PDF #00-021-1276, ICSD #62678) and $Ti_3O_5$ (PDF #01-072-2101, ICSD #20361) form in the HSC testing process (Figure 9). Also, well expressed highly intensive NaCl peaks (PDF #00-005-0565, ICSD #61662) are seen in the XRD curves.

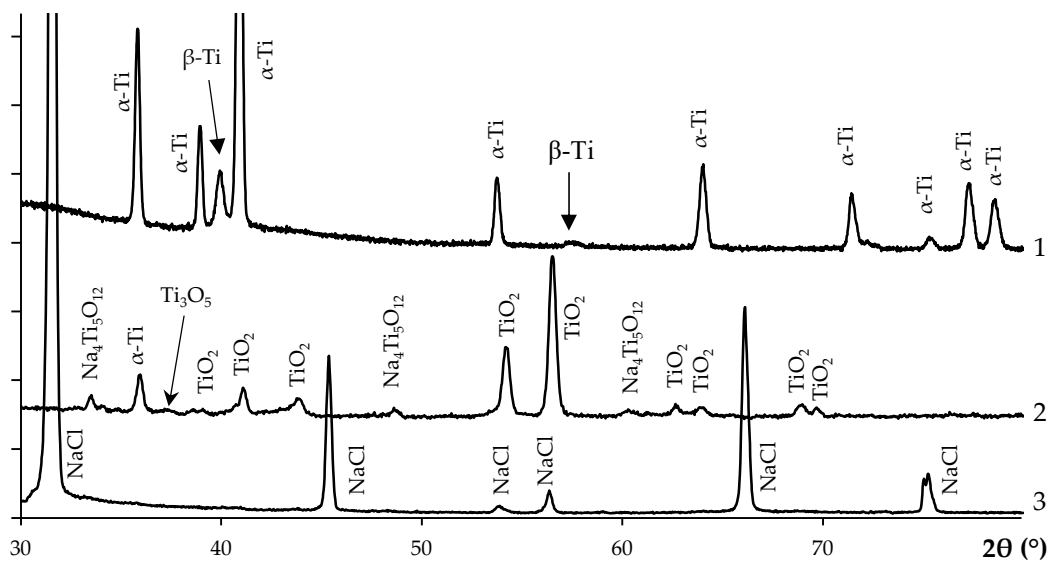

**Figure 9.** XRD spectra of the Ti6Al4V alloy specimen before (line (3)) and after rinsing (line (2)) in hot water. For comparison, the results of the XRD phase analysis of the Ti6Al4V alloys before HSC testing are presented in the figure (line (1)).

Note that the XRD peaks corresponding to NaCl and to $Na_4Ti_5O_{12}$ oxide disappeared after rinsing in hot water. It allows suggesting the $Na_4Ti_5O_{12}$ oxide to form on the external sides of the dark films whereas the Ti oxides $TiO_2$ and $Ti_3O_5$ at the inner sides of the ones adjacent to the Ti alloy specimens. This confirms the data [5] on the HSC of Ti alloys to be a multistage process and on a multilayered character of the resulting films indirectly [7].

The photographs of the polished side surfaces after HSC testing are presented in Figure 10. As can be seen from Figures 10a,d, an intensive uniform corrosion and decreasing of the specimen sizes were observed on the Ti (VT1-0) specimen' surfaces after HSC testing. No traces of local corrosion were observed. Also, traces of the intergranular corrosion (IGC) were observed on the surfaces of the Ti alloys Ti5Al2V and Ti6Al4V; general corrosion of the samples was almost absent (Figure 10). The mean depths of the corrosion defects in the Ti5Al2V and Ti6Al4V alloys were 40–50 μm and 20–40 μm, respectively. Note that the average depth of the corrosion defects in the Ti5Al2V alloy specimens obtained by SPS (40–50 μm) was considerably lower than in the coarse-grained Ti5Al2V alloy specimens in the initial state tested in the same HSC conditions (~400–600 μm [46]).

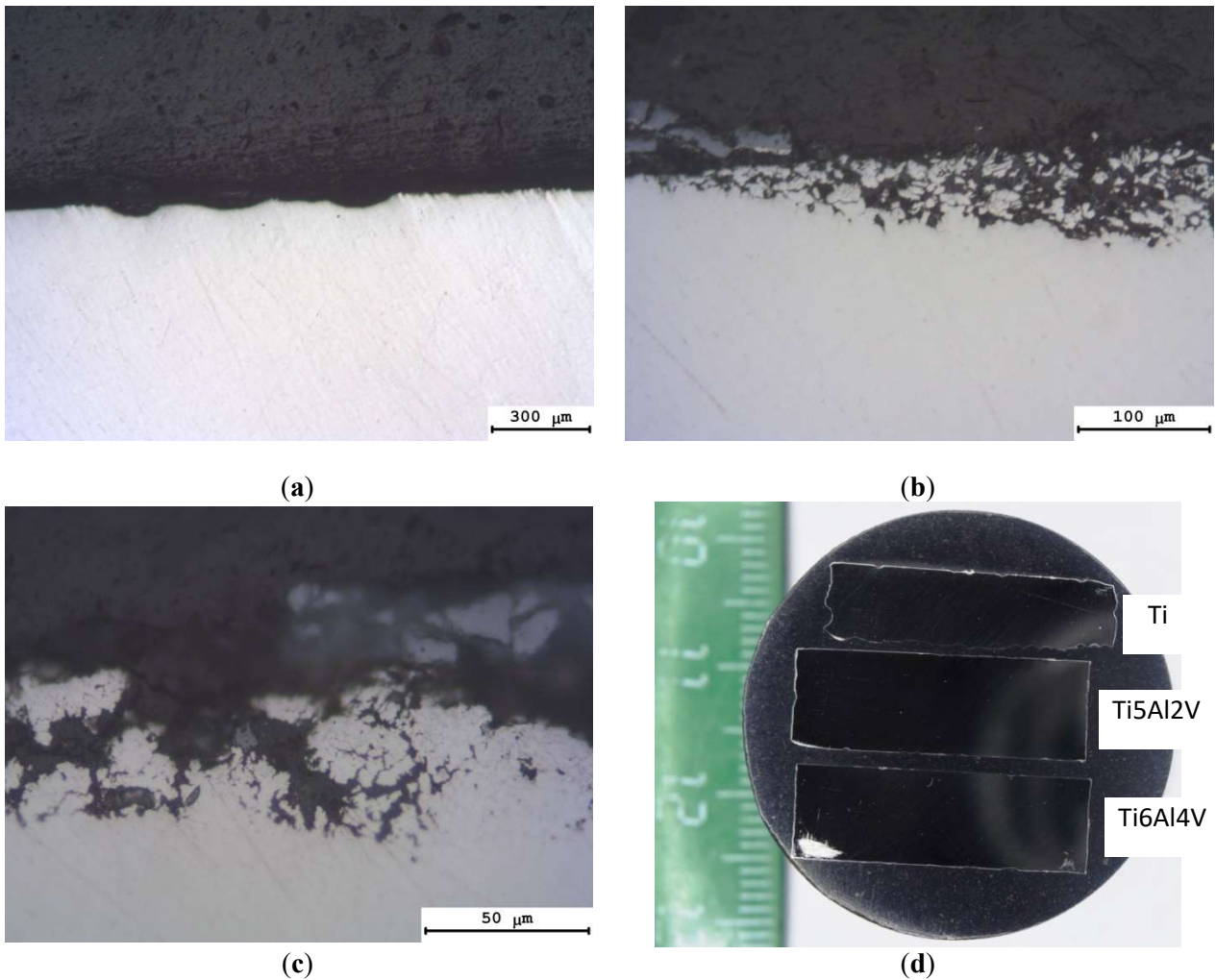

**Figure 10.** Optical microscopy images of the corrosion destruction character of the Ti alloys' specimens after HSC testing: (**a,d**) - Ti; (**b,d**) - Ti5Al2V alloy; (**c,d**) - Ti6Al4V alloy. Metallography.

This conclusion agrees qualitatively with the results of investigation of the specimens' mass changes after corrosion testing. These investigations have shown the mass of the Ti specimen to decrease by $\Delta m$ ~0.63 g whereas the one of the Ti6Al4V alloy by $\Delta m = 1.7 \cdot 10^{-3}$ g during 500 h of testing. The mass of the Ti5Al2V alloy specimen didn't change within the measurement uncertainty ($\pm$ 0.00035 g). The measurements of the specimens' masses were performed after rinsing in a hot water flow allowing removing readily soluble NaCl and $Na_4Ti_5O_{12}$ deposits. So far, the corrosion rates (the corrosion depth indices) for the specimens of the Ti (diameter 20.3 mm, height 5.5 mm) and of the Ti6Al4V one (diameter 20.4 mm, height 6.9 mm) were ~0.77 and ~$2.2 \cdot 10^{-3}$ mm/year respectively. So far, the specimens of the Ti in the HSC conditions had the 6th–7th grade of corrosion resistance caused mainly by intensive general (uniform) corrosion. The Ti6Al4V alloy had the 2nd grade of corrosion resistance whereas the Ti5Al2V alloy obtained by SPS had the 1st grade of the corrosion resistance. A higher corrosion rate of the Ti specimens as compared to the Ti5Al2V and Ti6Al4Vis observed in Figure 10d. As one can see in the figure, the Ti specimen had a considerably less thickness after HSC testing.

The results of the electrochemical investigations of the corrosion resistance of the Ti alloys obtained by SPS are presented in Figures 11 and 12. As can be seen from Figure 11 the passivation sages starting from the potential near –200 mV are manifested clearly in the potentiometric dependencies $i(E)$ of all investigated Ti alloys (Zone 1). The current densities were very low within these stages and were close to each other for all alloys. An increasing of the maximum current density $i_{max}$ at the passivation potential $E_{p(1)} \sim -370 \pm 10$ mV with increasing doping impurity contents was observed. The lowest values of $i_{max}$ ($i_{max(1)}$, $i_{max(2)}$) were observed for the VT1-0 material, the highest ones for the Ti5Al2V and Ti6Al4V alloys (Figure 11). The corrosion current densities of the reactivation curves $i_{max(2)}$ were ~2–2.5 times higher than the ones of the passivation curves $i_{max(1)}$.

Similar results were obtained in the electrochemical investigations of the corrosion resistance in Zone 2 (the edges) of the Ti alloys' specimens. Note that the values of the passivation potential $E_{p(2)}$ in Zone 2 were close to the ones of $E_{p(1)}$. However, the highest values of the corrosion current density in the reactivation curves $i_{max(2)}$ were lower essentially and were close to $i_{max(1)}$.

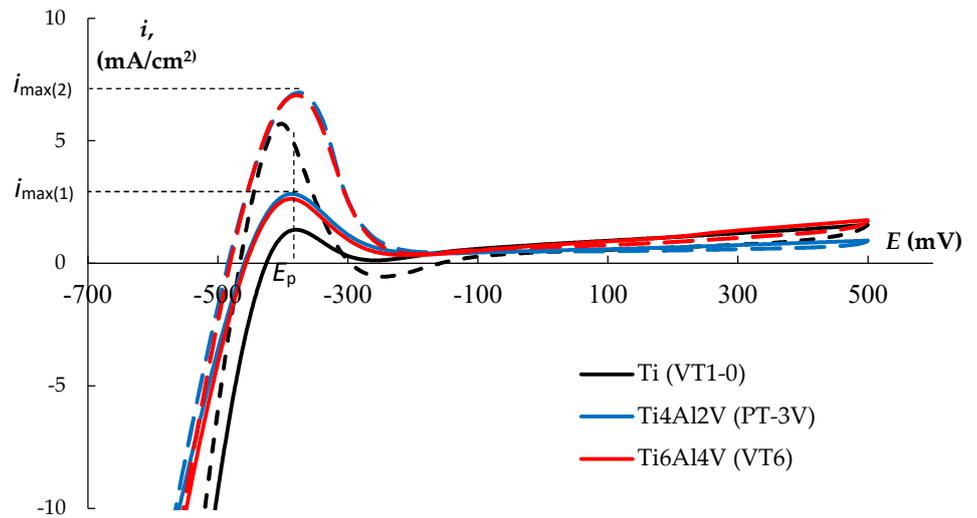

**Figure 11.** Potentiometric dependencies $i(E)$ for the Ti alloys' specimens obtained by SPS. Solid lines - ascending potential (passivation), dashed lines - descending potential (reactivation). Black curves - Ti, blue curves - Ti5Al2V alloy, red curves - Ti6Al4V alloy. Data for Zone 1 (central parts of the specimens).

As can be seen from Figure 12a, the values of the alloy potential in the chronopotentiodynamic testing in aqueous solution 0.2%HF + 10%HNO$_3$ reach the stationary values $E_{st}$ (which are presented in Figure 12b) rapidly. Note that the values of potential $E_{st}$ shift towards the negative region with increasing doping levels of the alloys and increasing volume fractions of the β-phase particles. Also, it is interesting to note that the values of $E_{st}$ for the central parts of the specimens of all alloys were more positive than for the ones on the sides of the specimens contacting the graphite mold directly.

Similar results were obtained by analysis of the dependencies lg$i(E)$. The results of electrochemical testing and analysis of the lg$i(E)$ curves are summarized in Table 3. As can be seen from Figure 12, the increasing of the doping level resulted in a shift of the corrosion potential $E_{cor}$ towards more negative region. Also, the materials of the central parts of the specimens had more positive values of the corrosion potential $E_{cor}$ than the ones of the specimens' edges (by ~10–20 mV).

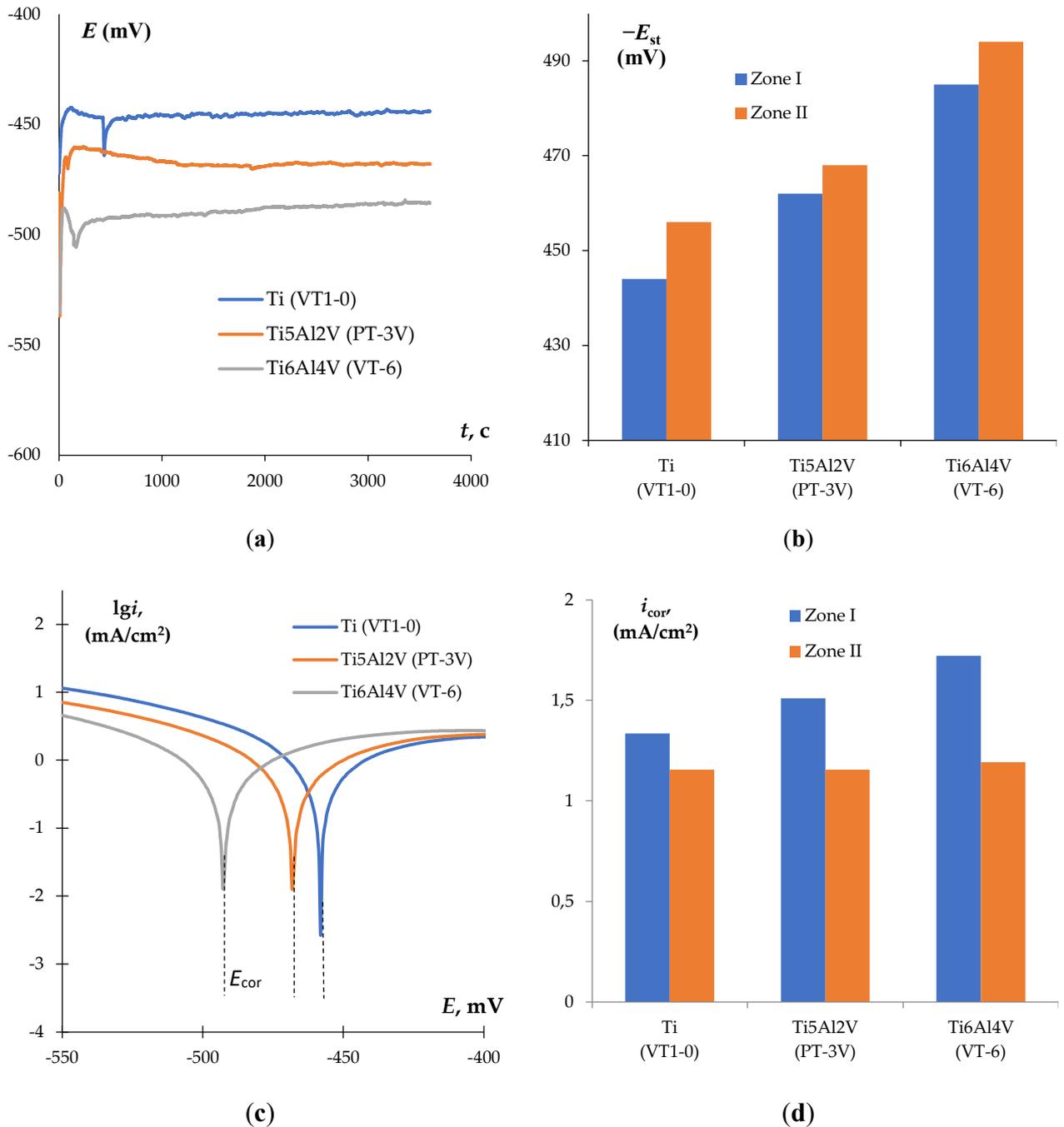

**Figure 12.** Results of electrochemical investigations of Ti alloys: (**a**)–dependence of potential on the time of exposure to the electrolyte $E(t)$ (Zone 2); (**b**) the values of stationary corrosion potential $E_{st}$ for the alloys; (**c**) Tafel curves $\lg(i)$-$E$ (Zone 2); (**d**) the values of the corrosion current density $i_{cor}$ for the Ti alloys

The analysis of the data presented in Figure 12d shows that the corrosion current density to increase with increasing doping level. This increasing was expressed the most clearly for the central parts of the specimens (Zone 1) while the changes in the corrosion current density in Zone 2 were not so significant. Note that the values of the corrosion current density for the alloy specimens obtained by SPS were comparable to the values of $i_{cor}$ for the coarse-grained alloy (~1.07 mA/cm$^2$) and for an

UFG Ti2.5Al2Zr (Russian industrial name PT-7M) alloy tested in the same medium (0.80–1.16 mA/cm$^2$) [15]. The corrosion current densities for the Ti5Al2V alloy specimens obtained by SPS were comparable to the values of $i_{cor}$ for the coarse-grained (~0.66 mA/cm$^2$) and UFG (~1.3 mA/cm$^2$) alloys Ti5Al2V after diffusion welding by SPS [51]. So far, the Ti alloy specimens obtained by SPS have high corrosion resistance comparable to the one of the industrial Ti alloys fabricated by traditional technology (hot deformation in the β-region with gradual decreasing of temperature from the β-region to the (α + β)-one).

**Table 3.** Results of electrochemical corrosion tests of the Ti alloy specimens.

| Alloy | Zone I | | | | | Zone II | | | | |
|---|---|---|---|---|---|---|---|---|---|---|
| | $E_{st}$, mV | $E_{cor}$, mV | $i_{cor}$, mA/cm$^2$ | A | C | $E_{st}$, mV | $E_{cor}$, mV | $i_{cor}$, mA/cm$^2$ | A | C |
| Ti (VT1-0) | −444 | −442 | 1.34 | 121 | 71 | −456 | −458 | 1.16 | 120 | 74 |
| Ti5Al2V (PT-3V) | −462 | −460 | 1.51 | 77 | 102 | −468 | −467 | 1.17 | 85 | 95 |
| Ti6Al4V (VT-6) | −485 | −484 | 1.72 | 78 | 100 | −494 | −492 | 1.19 | 83 | 97 |

The results of the samples after electrochemical tests surface studies indicate that the process of corrosion fracture has predominantly intergranular character (see Figure 13). In the two-phase α + β alloy Ti6Al4V a more intense corrosion destruction of the grain boundaries is observed compared with near-α alloy Ti5Al2V. This, in our opinion, also indirectly indicates that the parameters (size, volume fraction, nature of spatial distribution, chemical composition) of the β-phase particles have the most significant influence on the corrosion resistance of SPSed titanium alloys.

The hardness of the Ti alloys after the long-term (500 h) HSC testing at 250 °C remained very high. The mean hardness of the Ti was 1.9 GPa, the ones of the Ti5Al2V and Ti6Al4V alloys were 3.4 GPa and 3.3 GPa, respectively. No essential differences in the hardness values of the central parts of the specimens and aside were observed. The imprints from the diamond pyramidal indenter had a regular shape. The absence of cracks evidenced the sintered Ti alloy specimens to retain plasticity.

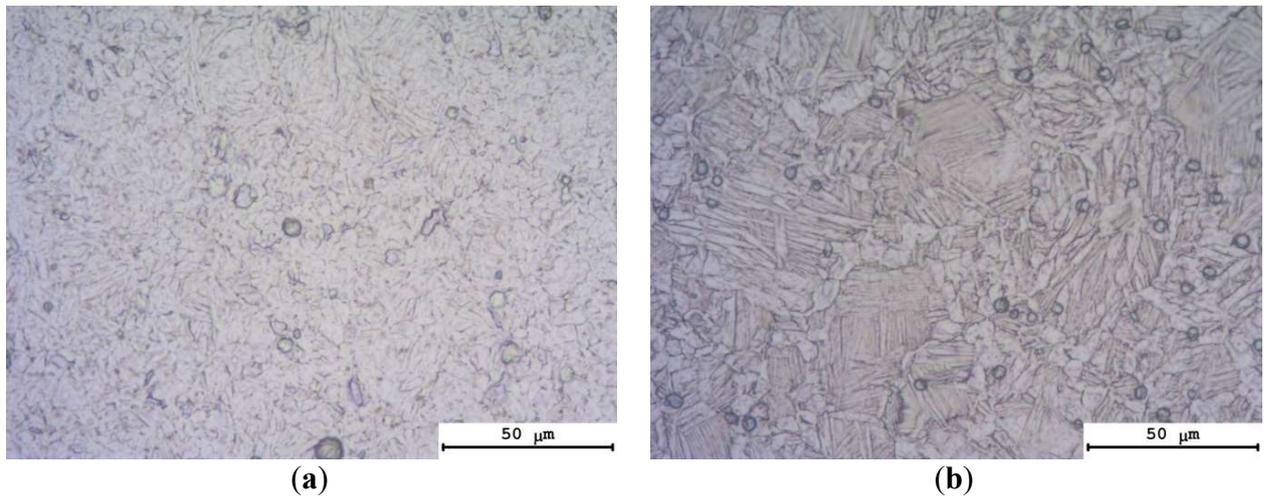

(a)                          (b)

**Figure 13.** Optical microscopy images of the specimens' surfaces of the Ti5Al2V (**a**) and Ti6Al4V (**b**) alloys after electrochemical testing. Metallography.

Summing up the results of the experimental investigations, we can conclude the specimens obtained by SPS have high hardness and corrosion resistance. However, significant differences in the corrosion properties of the central parts of the specimens and of the specimen edges were observed.

## 4. Discussion

SPS of the Ti specimens was performed in a graphite mold, through which millisecond high-power current pulses are passed. This ensures a rapid heating of the mold together with the specimen as well as prevents intensive plastic deformation of the Ti alloy leading to loosing of the specimen shape stability and to arising of tensile internal strain.

As has been shown earlier in a number of works [52–56], an intensive diffusion of carbon from the graphite mold into the surface layers of the sintered specimens in the SPS process is possible for some materials (Ni-W alloys [52,53], hard alloys based on tungsten carbide [54,55], garnet-based ceramics [56], etc.). It may lead to altering the phase composition and physical and mechanical properties of the surface layers (see, for example, [54,55]). (The possible effect of oxide film on the Ti alloy particle surfaces was neglected since an efficient dissociation of oxides at low enough SPS temperatures was reported [57]).

According to [58], the magnitude of the pre-exponential factor $D_{v0}$ for the diffusion coefficient of carbon $^{14}C$ in Ti is 3.02–3.18 cm$^2$/s in the temperature range 1223–1923 K (the range of existence of the β-phase). The activation energy $Q_v$ in the specified temperature range varies from 79 to 84 kJ/mol (~4.9–5.2 k$T_m$ where $T_m$ = 1943 K is the melting point of Ti). A numerical estimate of characteristic diffusion mass transfer scale $x_{diff}$ according to well-known formula $x_{diff} = (D_v \cdot t)^{1/2}$ at 850 °C (1123 K) and the holding time 10 min gives $x_{diff}$ ~ 2 µm. Probably, a greater depth of carbon penetration inside the Ti specimens is caused by the two factors. First, it is necessary to take into account the effect of a high gradient of the carbon concentration grad$C$ arising at the contact of the graphite mold walls with the specimen surface, which would lead to an increased carbon atoms flux according to Fick Equation: $I = -D_v \text{grad} C$. Second, it is worth noting that the diffusion of carbon takes place at the uniaxial stress applied, the magnitude of which (70 MPa) exceeds the yield strength at 850 °C. It leads to the creep mechanism to take place [39], which also promotes accelerated carbon diffusion into the Ti alloy specimens' surfaces.

It should be emphasized here that the effect of increased carbon concentration in the Ti specimens obtained by SPS was also noted in [59]. In [59] the carbon concentration in the Grade 2 purity Ti specimens after compaction by SPS was shown to increase from 0.009 to 0.048 wt.% without essential changes in the oxygen concentration. The authors of [59] didn't specify the parts of the specimens where the alloy compositions were analyzed in but pointed to an interaction of Ti with graphite at elevated temperature as a possible origin of increased carbon concentration. This hypothesis is supported indirectly by the differences in the crystal lattice parameters of α-Ti measured in the centers of the specimens (Zone 1) and aside (Zone 2). The investigations made in the "narrow slit" mode with the mean diameter of the X-ray beam ~2 mm have shown that the unit cell parameters $a$ and $c$ of α-Ti in the centers of the HCP specimens (Zone 1) to be lower than the ones in Zone 2 (at the edges of the specimens). In particular, the differences of the magnitudes $a$ and $c$ between the center and the edge of the VT1-0 specimen were 0.0020 Å and 0.0028 Å, respectively. This corresponds to a change in the α-Ti unit cell volume by 0.0686 Å$^3$. The authors consider it necessary to note that although the changes in the crystal lattice parameters were observed reliably but exceed the measurement uncertainty only slightly.

In our opinion, the effect of carbon diffusion into Ti allows explaining a number of observations made.

First, as noted above, the values of hardness of the alloys obtained by SPS were rather high. In particular, the hardness of the Ti5Al2V alloy obtained by SPS was close to the one of the UFG Ti5Al2V alloy with the grain sizes ~0.5 μm obtained by Equal Channel Angular Pressing (~3.1–3.2 GPa) [39,46] whereas the hardness of the as-delivered Ti5Al2V alloy didn't exceed 2.0–2.1 GPa. We suppose, it may also be related to the intensive diffusion of carbon from the graphite mold surface inside the specimens. As it has been shown in [60], doping of Ti with carbon (up to 0.2 wt.%) may lead to a considerable increasing of strength of Ti both in deformed state and in the recrystallized one.

It is worth noting that, in our opinion, a considerable part of excess carbon in Ti is in a supersaturated solid solution state, not in the form of Ti carbide Ti-C particles. The intensive carbon diffusion into the surface layers of the specimens takes place at 850 °C whereas rapid cooling down of the specimens after 10-min holding allows freezing a considerable fraction of the "excess" carbon inside the α-Ti crystal lattice. It is pointed to indirectly by the shifts of the stationary potential $E_{st}$ and of the corrosion potential $E_{cor}$ measured near the specimens' edges towards the negative region as compared to the ones measured in the central parts of the Ti alloy specimens. This shift originates from the distortion of the α-Ti crystal lattice due to the formation of the supersaturated solid solution of carbon. Note also that the positive effect of carbon doping on the corrosion resistance of the Ti near-α alloy was reported in [61]. The effect of increased corrosion resistance of Ti by carbon doping was employed in patent RU 2418086 C2 to develop a new Ti alloy.

So far, the effect of the diffusion saturation of the Ti alloy surfaces with carbon allows explaining the increased hardness and corrosion resistance of the Ti alloys observed in the experiment.

Finally, the origin of high resistance of the sintered Ti alloys to HSC should be discussed. In [15,46], the resistance of the Ti-Al-V alloys to HSC was shown to depend on the concentrations of the corrosion-aggressive elements (first of all vanadium) at the Ti grain boundaries as well as on the volume fraction of the β-phase particles. The β-phase particles contain an increased concentration of doping elements stabilizing the β-phase (V in the Ti-Al-V alloys). Therefore, the presence of the β-phase particles also results in the accelerated corrosion destruction of the interphase α/β-boundaries. As shown above, the average grain sizes in the Ti alloys obtained by high-speed sintering were close to the ones of the initial particles. We can suggest the diffusion redistribution of the doping elements inside the sintered specimens and, hence, the formation of segregations at the grain boundaries as well as the nucleation of the β-particles to take place in the SPS process (the sintering was performed in the intermediate α + β region). It leads to formation of the places of accelerated destruction in the Ti alloys via the chemical and/or electrochemical corrosion mechanism. Consequently, the corrosion mechanism changes from the uniform corrosion in Ti to the intergranular corrosion in the Ti-Al-V alloys. It can be assumed that the fraction of the nucleated β-phase particles is low due to high heating rates whereas the V concentration at the Ti grain boundaries appears to be much lower than the coarse-grained Ti alloys. This provides high resistance of the Ti alloys obtained to IGC in the HSC testing conditions as well as in the electrochemical testing.

## 5. Conclusions

The specimens of Ti alloys: VT1-0 (α-alloy), Ti5Al2V (near-α-alloy) and Ti6Al4V (two-phase α + β alloy) were produced by Spark Plasma Sintering. The alloys have high density (close to the theoretical one) and high hardness, obviously due to diffusion of carbon from the graphite mold surface into the specimen one.

All studied Ti alloy specimens obtained by SPS had high strength and good plasticity in tension testing at room temperature. The alloy specimens obtained by SPS have high corrosion resistance in the electrochemical testing as well as in long-term high-temperature testing in a mixture of crystalline salts. It was shown that the VT1-0 Ti specimens are destroyed in the hot salt corrosion conditions by the general corrosion mechanism whereas the specimens of the Ti alloys of the Ti-Al-V system (Ti5Al2V and Ti6Al4V) by the intergranular corrosion mechanism. The depths of the corrosion defects in the Ti5Al2V alloy produced by SPS were less than the ones in the specimens of as-delivered coarse-grained Ti5Al2V alloy by an order of magnitude.

The obtained results demonstrate the wide prospects of application of SPS technology for the modification of the structure of the Ti alloys in order to improve the physical and mechanical properties and operational characteristics of the ones further. It is important to note that the application of SPS technology allows providing a high level of mechanical properties and corrosion resistance of the Ti alloys without application of intensive plastic deformation technology leading to the grinding of the grain structure (see [15,46]). It is especially important in the case of further exploitation of the products from the Ti alloys at elevated temperatures and stresses when the creep rate in the fine-grained materials may increase considerably and lead to change of the geometry of the highly responsible products and constructions.